\documentclass[10pt,twocolumn]{article}
\pdfoutput=1

\usepackage[hidelinks]{hyperref}
\hypersetup{
	bookmarks=true,         
	unicode=false,          
	pdftoolbar=true,        
	pdfmenubar=true,        
	pdffitwindow=false,     
	pdftitle={Autonomous Reconfiguration Procedures for EJB-based Enterprise Applications},
	pdfauthor={Thomas Vogel, Jens Bruhn, and Guido Wirtz},
	pdfsubject={},
	pdfcreator={},
	pdfproducer={},
	pdfkeywords={maintenance, seamless reconfiguration, EJB},
	pdfnewwindow=true,
}

\usepackage{latex8}
\usepackage{times}
\usepackage{graphicx}
\usepackage{url}
\usepackage{array}
\begin{document}

\title{~\\[-1.5cm] Autonomous Reconfiguration Procedures for EJB-based Enterprise Applications\\[-0.5cm]}
\author{Thomas Vogel, Jens Bruhn, and Guido Wirtz\\
Distributed and Mobile Systems Group, University of Bamberg\\
Feldkirchenstra\ss e 21, 96052 Bamberg, Germany\\
th.vogel@gmail.com,
$\{$jens.bruhn$|$guido.wirtz$\}$@uni-bamberg.de\\[-0.5cm]
}

\maketitle

\begin{abstract}
\noindent
Enterprise Applications (EA) are complex software systems for supporting the business of companies. Evolution of an EA should not affect its availability, e.g., because of a temporal shutdown, business operations may be affected. One possibility to address this problem is the seamless reconfiguration of the affected EA, i.e., applying the relevant changes while the system is running. Our approach to seamless reconfiguration focuses on component-oriented EAs. It is based on the Autonomic Computing infrastructure \emph{mKernel} that enables the management of EAs that are realized using Enterprise Java Beans (EJB) 3.0 technology. In contrast to other approaches that provide no or only limited reconfiguration facilities, our approach consists of a comprehensive set of steps, that perform fine-grained reconfiguration tasks. These steps can be combined into generic and autonomous reconfiguration procedures for EJB-based EAs. The procedures are not limited to a certain reconfiguration strategy. Instead, our approach provides several reusable strategies and is extensible w.r.t. the opportunity to integrate new ones.
\\[4pt]
\end{abstract}
\begin{Section}{Introduction}
\label{sec:intro}
\noindent
\emph{Enterprise Applications} (EA) are complex software systems for supporting the business of a company. According to Lehman's laws \cite{637156} software 	implementing real world applications like EAs must continually evolve, else 	their use and value would decline. The need for system evolution originates, 	e.g., from failures, inefficiencies or changes of the business or of the system environment that lead to new or changing requirements for EAs. Thus, system evolution can be categorized as \emph{corrective} (removing software faults), \emph{adaptive} (adjusting the system to the changing environment), or \emph{perfective} (enhancing or improving the functional and non-functional system characteristics) (cf. \cite{302181,807723}). Due to the critical role of an EA within a company this evolution should not affect the availability of an EA and therefore the business operations. Otherwise, the company might miss business opportunities and loose reputation and trust. One approach to address this problem is the \emph{seamless} reconfiguration, i.e., applying the relevant changes to the system while it is running. Except of delays in the response time reconfiguration should be transparent to the clients of the EA. This \emph{post-deployment runtime evolution} can be seen as one critical challenge in software evolution \cite{1572302}. 
To cope with this issue, the modularity of software systems, as proposed by the concept of \emph{Component Orientation} (CO) \cite{515228}, and the automation of system maintenance tasks, as described by the vision of \emph{Autonomic Computing} (AC) \cite{1014770, 642200}, can help. With the \emph{mKernel} system \cite{aiccsa, mkernel} a generic AC infrastructure is available that enables 	comprehensive management of component-oriented EAs that are realized with the 	\emph{Enterprise Java Beans (EJB) 3.0} technology \cite{jsr220}. Based on 	\emph{mKernel}, we provide a comprehensive set of steps, that are customizable 	and perform fine-grained reconfiguration tasks. These steps can be combined 	flexibly to generic and autonomous reconfiguration procedures for EJB-based EAs. Each of these procedures realizes a certain \emph{reconfiguration strategy}, i.e., a certain way to perform a reconfiguration. Currently, our approach provides four reusable strategies that serve as templates for easing the planning and execution of a concrete reconfiguration.
	
The rest of the paper is structured as follows: Section \ref{sec:background} provides an overview to the background, namely system reconfiguration, CO and the AC infrastructure \emph{mKernel}. Section \ref{sec:relatedwork} discusses related work, while section \ref{sec:reconf} presents our approach of reconfiguration procedures. Finally, the last section gives a conclusion and an outlook on future work.
\end{Section}

\begin{Section}{Background}
\label{sec:background}
\noindent
After introducing the basics of system reconfiguration, the concept of CO and relevant aspects of the EJB standard are presented. Finally, we describe how \emph{mKernel} combines EJB with the vision of AC.
	
\begin{SubSection}{System Reconfiguration}
\label{subsec:reconf}
\noindent
The architecture of a software system is the \emph{high-level organization of [its constituent] computational elements and the interactions between those elements} (\cite{10.1109/TSE.1995.10003}, p.269). In this context, according to \cite{10.1109/MC.2004.48}, there are two general approaches for software reconfiguration: \emph{parameter adaptation} and \emph{compositional adaptation}. The first one modifies variables of one or more elements that determine their behavior. The second one addresses structural reconfiguration through addition and removal of elements, including the manipulation of connections among them (cf. e.g. \cite{60317, 302181, 760388}). The weakness of parameter adaptation is that it allows only changes that were anticipated during development, because the elements have to provide the variables and react appropriately to their modifications. In contrast, compositional adaptation is intended for the dynamic and unanticipated reconfiguration of a system. 

For carrying out a reconfiguration, two objectives are to be considered and desirable \cite{60317}. First, the reconfiguration should minimize the disruption to the system, i.e., the affected part of the system may notice delays but no failures, while the rest of the system should be able to continue its execution normally. Thus, reconfiguration should be carried out \emph{seamlessly}. Second, a consistent state of the system must be preserved during and after reconfiguration. Consequently, a reconfiguration, like, e.g., the replacement of an element, may require to place the affected part of the system in a consistent state before structural changes are performed. A state is consistent if the affected elements are \emph{quiescent} \cite{60317}, i.e., none of them is currently engaged in servicing a request and none of them will initiate a request. Furthermore, no requests initiated by non-affected elements are forwarded to affected ones.
To reach a quiescent state, requests that are currently serviced must be 		finished. New requests must be blocked except those which are needed to finish servicing ones. Otherwise, some elements are not able to reach a quiescent 		state and end up in a deadlock. Quiescence of the affected part of the system gives new elements the opportunity to be initialized in a state which is consistent with the rest of the system, and elements to be removed the 		opportunity to leave the system in a consistent state \cite{60317}. In case of an element replacement, this may include the need for transferring the internal state of a replaced element to a replacing one \cite{302181, rosaAuto}.
		
How to apply changes are questions of reconfiguration strategies. In \cite{rosaAuto} the three strategies \emph{Flash}, \emph{Non-Interrupt} and \emph{Interrupt} are presented. The \emph{Flash} strategy reconfigures one element without concerning about other elements. Reconfiguration takes place immediately without handling existing interactions and the states of the affected elements. No state transfer is performed and existing connections to old elements are not updated. Therefore, these connections become invalid and are likely to cause errors. Finally, the system may become inconsistent. 
Consequently, \emph{Flash} does not always perform a seamless and consistent reconfiguration. Nevertheless, it can be used, amongst others, for parameter adaptation or for reconfiguring elements not being critical for the consistency of the application. In contrast, the other strategies preserve consistency of the system and perform a seamless reconfiguration. 
The \emph{Non-Interrupt} strategy supports the exchange of elements without the need for quiescence, hence reducing system disruption significantly. Both elements, the old one that is going to be replaced and the replacing one, are active. An intercepting facility forwards requests of already existing sessions to the old one and requests of new sessions to the	replacing one. After all sessions on the old element have finished, it can be removed and only the new element is used. This strategy does not require a state transfer. It requires that the two elements can be used concurrently.
The \emph{Interrupt} strategy transfers the affected part of the system into a quiescent state before reconfiguration takes place. The states of the affected elements and existing connections between elements are handled, such that, e.g., an element replacement can be performed without causing any failures. After reconfiguration, the affected part of the system is released at once from the quiescent state, such that it can be assured that all elements and connections are reconfigured appropriately, before resuming their execution. Comparably with the other strategies, an advantage of requiring quiescence is that, e.g., an underlying database is not used during quiescence, which enables its consistent modification or transfer. 
Consideration of strategies is important to find the best way to reconfigure a system.
\end{SubSection}

\begin{SubSection}{Component Orientation}
\label{subsec:co}
\noindent
The concept of \emph{Component Orientation} (CO) \cite{515228} is a paradigm for the development of software systems in a modular way through functional decomposition. Such systems are composed of modules, called \emph{Components}. A component encapsulates a certain functionality and provides it through contractually specified \emph{Interfaces}. A component may use services from other components through their provided interfaces. An interface required by a component is called \emph{Receptacle}. Consequently, a component-based system can	be seen as a collection of loosely-coupled modules which collaborate among each other through their interfaces. Furthermore, a component can be deployed independently and is subject to composition by third parties \cite{515228}. Thus, CO addresses the complexity during development and deployment by modularity of requirements, architectures, designs, implementations and deployments. This modularity supports the partial reconfiguration of component oriented systems.
		
The \emph{Enterprise Java Beans} standard (EJB), version 3.0, \cite{jsr220}	is a component standard for the realization of component-oriented EAs on top of the \emph{Java} programming language. It defines a sound component model that is based on so called \emph{Enterprise Beans}, or \emph{Beans} for short. There are two types of beans considered in the standard, namely \emph{Message Driven Bean} and \emph{Session Bean}. The former one is intended to be accessed through asynchronous message passing, and the latter one provides interfaces to	access its encapsulated functionality. Session beans can be either	\emph{stateless} or \emph{stateful}. An instance of a stateful session bean is exclusively used by a single client and retains its client-specific \emph{Conversational State} across multiple invocations. In contrast, an instance of a stateless session bean is not exclusively used by a client. Moreover, each invocation from a client on the same reference may be executed on different instances. Thus, all instances of one stateless session bean are equivalent, and their states are client independent.
Receptacles can be declared for session and message driven beans through \emph{EJB References}. These can be connected to interfaces provided by session	beans. Beans may be customized through \emph{Simple Environment Entries} which can be interpreted as a kind of property. Before deploying beans, they must be configured, i.e., their properties must be set and their corresponding receptacles and interfaces must be connected. As unit of deployment the EJB standard defines the \emph{EJB module} that must contain at least one bean. In the EJB context, parameter adaptation is performed through setting bean properties, and compositional adaptation through (un)deploying modules and manipulating connections between beans. However, after the deployment of a module into an \emph{EJB Container}, the runtime environment for components, the configurations of beans can not be changed.
\end{SubSection}

\begin{SubSection}{Autonomic Computing and mKernel}
\label{subsec:ac}
\noindent
The vision of \emph{Autonomic Computing} (AC) \cite{1014770, 642200} addresses the management of systems at runtime. Its basic idea is to assign low level, administrative tasks to the managed system itself to disburden human administrators. The system manages itself according to the goals specified by the administrator. Autonomous management covers the four aspects \emph{self-healing}, \emph{self-protection}, \emph{self-optimization}, and \emph{self-configuration}. The last aspect addresses reconfiguration explicitly. Furthermore, each aspect can be mapped to at least one of the different kinds of system evolution discussed in section \ref{sec:intro}, namely corrective, adaptive, and perfective.

The \emph{mKernel} system \cite{aiccsa, mkernel} provides a generic AC infrastructure for EJB-based autonomous applications. It includes a comprehensive \emph{Application Programming Interface} (API) of \emph{sensors} and \emph{effectors} through which the managed application can be inspected and manipulated by a higher level facility. Through this API, \emph{mKernel} provides a reflective view, the \emph{meta level}, of the managed application, the \emph{base level}. Both levels are causally connected \cite{38821}. This reflective view enables the management of the application at three different levels of abstraction. 
The \emph{Type Level} addresses information regarding types of the constituting elements of the managed application, i.e., artifacts being the result of development. 
The \emph{Deployment Level} concentrates on a concrete configuration of the managed application that is deployed into a container. 
Finally, the \emph{Instance Level} addresses the bean instances and interactions among them. With this multi-level view, subtle management operations are possible. 
As discussed in section \ref{subsec:co}, the EJB standard limits the configuration of bean properties and connections among beans to the deployment time, but \emph{mKernel} enables the modifications of them at runtime. Together with supporting the lifecycle of EJB modules, \emph{mKernel} provides runtime support for parameter and compositional adaptation. Nevertheless, the EJB specification is not violated or restricted by \emph{mKernel}. Developers of EJB modules do not have to follow special guidelines beyond those of the EJB standard during development to enable the manageability of modules through \emph{mKernel}. Thus, the developer can solely focus on the application logic while a preprocessing tool weaves the sensors and effectors into the EJB module. This approach maintains the idea of \emph{Separation of Concerns}.
Furthermore, \emph{mKernel} is realized as a plugin for an existing EJB container and does not require any adjustments of the container implementation. 
\end{SubSection}
\end{Section}

\begin{Section}{Related Work}
\label{sec:relatedwork}
\noindent
Our approach to seamless reconfiguration is inspired by the work of Rosa, Rodrigues and Lopes \cite{rosaAuto} who present a framework for message-oriented systems that supports a fixed set of reconfiguration strategies. In contrast to their work, our approach is extensible w.r.t. the integration of new strategies. Moreover, the replacement of strategy elements is supported which provides additional flexibility. We support separation of concerns, because developers of EAs do not have to consider reconfigurations during development. 
Finally, the deployment and instance level are explicitly addressed, especially the transfer of conversational states of stateful session bean instances is supported. Our work addresses a different application area, namely EJB-based EAs. 

G\"obel and Nestler \cite{984812} extend the EJB specification by adding one more bean type, namely a composite bean. This composite encapsulates runtime adaptation by selecting different sub-components of the composite. Thus, the developer must consider this extension to the standard and only anticipated reconfiguration is possible that depends on the internals of the composite.
Jarir, David, and Ledoux \cite{obasco:jarir-david-ledoux.wcop-ecoop2002} enhance the EJB container to provide limited reconfiguration by intercepting calls to impose user-defined functionality. More possibilities are provided by Rutherford et al. \cite{760388}, though their work is restricted to reconfiguration at the deployment level. They consider the management of the deployment lifecycle of modules and the modification of properties and of connections of beans. Nothing is said about the handling of bean instances, i.e., replacing bean instances together with their possible conversational states is not considered.
In contrast, Matevska-Meyer, Olliges, and Hasselbring \cite{matevskameyer-2004-DECOR04}, who confine reconfiguration to redeploying modules, recognize the problem of the state transfer. They conclude that stateful beans are not safe to structural changes and provide no solution. 
Finally, the research group of the \emph{Peking University Application Server} (PKUAS) \cite{1167760} has implemented an own EJB container that incorporates the necessary technological facilities for updating modules including bean instances and state transfer. Thus, they consider the deployment and instance level. But they do not support higher-level facilities, like reconfiguration strategies that may simplify the role of administrators.
\end{Section}

\begin{Section}{Autonomous Reconfiguration Procedures}
\label{sec:reconf}
\noindent
Our approach to seamless reconfiguration of EJB-based EAs  covers parameter and compositional adaptation. To meet various reconfiguration needs we identified and	provide a comprehensive and complete set of customizable and reusable \emph{steps}, that are described in table \ref{tab:table}. The first column contains identifiers for the steps. The second column covers a short description of the particular step. Each step performs a special reconfiguration task, like, e.g., the deployment of a module (step $a$) or the establishment of connections between beans (step $l$). Steps are realized by so called \emph{executors} that are based on the \emph{mKernel} API. This is depicted on the left hand side in the reconfiguration model in figure \ref{fig:reconf-model}. 
\begin{table*}[t]
\normalsize
\begin{tabular}[c]{|m{0.2cm}|m{12.3cm}|m{0.8cm}|m{0.3cm}|m{0.4cm}|m{0.3cm}|m{0.6cm}|}
\hline
\textbf{ID} & \textbf{Step} & \textbf{dep.} & \textbf{$F$} & \textbf{$NI$} & \textbf{$I$} & \textbf{$I/NI$}  \\ \hline \hline
$a$ & Deployment of the new EJB module. Setting the \emph{Simple Environment Entries} and connecting the \emph{EJB Reference} of its beans. & / & 1 & 1 &	1 & 1 \\ \hline
$b$ & Declaration of the quiescence region which comprises those beans or whole modules that must be quiescent at a later point in time. For module replacement, this region is the module which is going to be replaced. & / & - & - & 2 & 2 \\ \hline
$c$ & Start tracking and collecting session bean instances of beans of the quiescence region to get to know the instances that are handled with step $f$. & $b$ & - & - & 3 & 3 \\ \hline
$d$ & Initializing the quiescence, i.e., initializing the blocking of calls on the instances of beans of the quiescence region. The region becomes quiescent after finishing current calls. & $b$ & - & - & 4 & 6 \\ \hline
$e$ & Waiting until the quiescence region becomes quiescent. & $d$ & - & - & 5 & 7 \\ \hline
$f$ & Extracting the state of stateful session bean instances being collected because of step $c$. & $c$, $e$ & - & - & 6 & 8 \\ \hline
$g$ & Extracting the database that underlies the quiescence region. & $e$, $f$ & - & - & 7 & - \\ \hline
$h$ & Transfer or modify the database. & $g$ & - & - & 8 & - \\ \hline
$i$ & Starting of the new EJB module. & $a$ & 2 & 2 & 9 & 4 \\ \hline
$j$ & Modifying (optional) and injecting the states, being extracted at step $f$, to newly created instances of the corresponding stateful session beans of the new EJB module. & $f$, $h$, $i$ & - & - & 10 & 9 \\ \hline
$k$ & Publishing the references of the new bean instances that have been the target of the transfer of step $j$. Client components holding references to replaced stateful session bean instances are provided with the corresponding new reference to the replacing instance. & $j$ & - & - & 11 & 10 \\ \hline
$l$ & Re-route connections that are newly established. The source of the these connections are client components of the new/old EJB module and the target of these connections shifts from the beans of the old EJB module to the beans of the new EJB module. & $h$, $i$ & 3 & 3 & 12 & 5 \\ \hline
$m$ & Re-route already existing connections, i.e., clients of the old module holding references to bean instances of the old module are provided with new references to bean instances of the new module. In case of an $I$ or $I/NI$ this step only considers connections whose targets are bean instances which have not been transferred. Connections to transferred bean instances are already covered by step $k$. In case of $NI$ this step is optional and only consistently applicable if the target of the connection is a stateless session bean instance. & $h$, $i$ & - & 4 & 13 & 11 \\ \hline
$n$ & Release the quiescence region, i.e., blocked calls and eventually blocked bean instance lookup requests are released and continue executing through using the reference already held before quiescence or the reference provided in steps $k$, $l$ or $m$. & $d$, $h$, $k$, $l$, $m$ & - & - & 14 & 12 \\ \hline
$o$ & Stop and optionally undeploy the old EJB module if the old module is not used any more, or in case of the $F$ strategy, should not be used any more through existing connections. & $g$, $k$, $l$, $m$ & 4 & 5 & 15 & 13	\\ \hline
\end{tabular}
\caption{Reconfiguration procedures and their steps}
\label{tab:table}
\end{table*}
Our implementation provides default executors for all steps, except the step that is intended for reconfiguration of databases. Nevertheless, administrators have the freedom to provide \emph{custom executors} for arbitrary steps replacing the default ones. In this way, special requirements for reconfiguring concrete applications can be fulfilled.
	
\begin{figure}[htb]
	\vspace{-0.2cm}
	\centering
	\includegraphics[width=7cm]{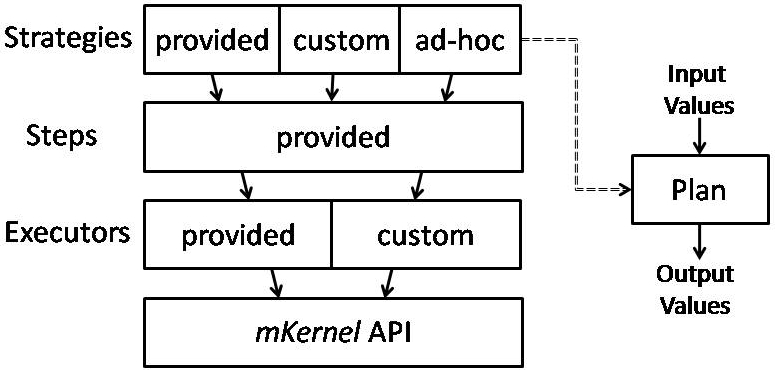}
	\caption{Reconfiguration Model}
	\label{fig:reconf-model}
	\vspace{-0.5cm}
\end{figure}
	
The provided steps are the basis for the \emph{strategies} (see figure \ref{fig:reconf-model}). Therefore, steps can be combined into generic and autonomous reconfiguration procedures. A procedure must fulfill the dependencies between its constituting steps. The third column of table \ref{tab:table} contains the steps each step depends on transitively. A '/' depicts that the particular step does not depend on any other step. Starting a new EJB module, e.g., requires that the module has been deployed before, therefore step $i$ depends on $a$. Nevertheless, as some steps may be optional, corresponding dependencies need not to be met. For the case, that no state transfer is necessary, steps $c$, $f$, $j$ and $k$ can be omitted, and the step of stopping the old EJB module ($o$) does not depend on step $k$, but only on $g$, $l$ and $m$. Consequently, these dependencies are influenced by a concrete arrangement of steps that may skip optional steps and by the concrete modules and beans each step is addressing. However, dependencies can be used to find basic restrictions in ordering the steps or potentials for parallel execution of steps. There exists, e.g., no dependency between the steps $a$ and $b$, such that they can be executed in arbitrary order or even in parallel. The reusability of each step, the flexibility in ordering the steps and the possibility to omit optional steps enable various combinations of steps into generic reconfiguration procedures. Each procedure realizes a certain reusable reconfiguration strategy. Therefore, administrators can develop \emph{custom} strategies, that may be derived from others or that may be completely new ones. Even, a dynamic arrangement of steps during runtime is possible, resulting in \emph{ad-hoc} strategies (see figure \ref{fig:reconf-model}).
Our approach provides the four strategies \emph{Flash} ($F$), \emph{Non-Interrupt} ($NI$), \emph{Interrupt} ($I$), and \emph{Interrupt/Non-Interrupt} ($I/NI$). Besides the first three ones, already presented in section \ref{subsec:reconf}, we identified $I/NI$ as an additional new strategy for replacing modules. It is a mixture of the strategies $I$ and $NI$. Its idea is that the new and old module are running concurrently, but newly created sessions are forwarded to the new module and start processing immediately. Already running sessions on the old module will not run until they finish, like it is done with the $NI$ strategy. In contrast, they are driven into a quiescent state and their instances of the stateful session beans are transferred to the new module, where finally the sessions continue their processing. The advantage of $I/NI$ is that system disruption is minimized because newly created sessions are not blocked from servicing requests. Therefore, the underlying database must be usable by both modules concurrently. Furthermore, the old module is removed from the system consistently.
	
In the following, we discuss in detail how the four strategies $F$, $NI$, $I$ and $I/NI$ can be applied for the case of replacing an EJB module with an alternative implementation.
Therefore, each of the last four columns of table \ref{tab:table} describes a realizing procedure for the corresponding strategy. The entries of these columns are to be read as follows. A step that is not applicable or available within a strategy is denoted with a '-'. Otherwise, the number indicates the position of this step within the procedure. 
	
For the $F$ strategy only the deployment level is relevant since existing bean instances and connections among them are not handled. The other strategies distinguish between already existing connections and newly established connections of bean instances, hence considering the instance level of the application. Re-routing a connection before it is created is always possible. Re-routing existing connections is feasible if the target of the connection is a stateless session bean, because the states of stateless session bean instances are client independent (see section \ref{subsec:co}) and both beans, replaced and replacing one, provide the same functionality. However, if the target is a stateful session bean, an existing connection can only be modified consistently if the conversational state is transferred to the corresponding target instance, otherwise the client-specific state would get lost. As described in section \ref{subsec:reconf} a state transfer requires quiescence of the affected beans, i.e., \emph{all} instances of the affected beans must be quiescent. Reaching quiescence is simplified by the EJB standard because bean instances are per definition non-reentrant and are not allowed to perform any kind of thread handling. Quiescence is performed by the steps $b$, $d$, $e$ and $n$. Another motivation for quiescence is the need to transfer or modify the database (steps $g$ and $h$) that underlies the modules, i.e., the old and the new module must be either quiescent or in a stopped state. This is addressed by the $I$ strategy. In summary, a state transfer (steps $c$, $f$, $j$ and $k$) is only required if stateful session beans are involved and an $I$ or $I/NI$ strategy should be used. Without state transfer and database reconfiguration there is no need for quiescence, hence $F$ or $NI$ are the preferable strategies.
	
For each step being part of a concrete strategy an executor must be assigned. A step executor may define input and output parameters. Inputs represent information required for an appropriate execution and information about execution results are provided through outputs. Outputs can be mapped to inputs of subsequent executors. E.g., our executor implementation for step $f$ outputs the extracted conversational states which are used as inputs for the executor of step $j$. At strategy level, inputs can also be specified. These can be connected to those executors inputs for which no matching outputs are given. Likewise, outputs can be defined for a strategy that provide information about execution results of an instantiated strategy to an administrator. Therefore, executor outputs can be connected to strategy outputs. To sum up, a concrete strategy consists of a set of steps together with their executors, specifications of inputs and outputs at the strategy level, and mapping specifications between parameters. In addition to the dependencies described in the third column of table \ref{tab:table}, these mappings may	introduce additional dependencies. A strategy is \emph{valid} if there are no circular dependencies and if all executor inputs are connected either to strategy inputs or outputs of preceding executors. As long as the dependencies are fulfilled, the order of steps may change within a procedure. Thus, it is conceivable that a strategy is realized by several procedures, i.e., different orders of steps. The procedures described in the last four columns of table \ref{tab:table} reflect the provided implementations. For a concrete reconfiguration need, an administrator must provide a reconfiguration plan, i.e., a strategy must be chosen, instantiated and configured. Consequently, the plan consists only of the selected strategy and of values for strategy inputs (see right hand side of the figure \ref{fig:reconf-model}). During execution, parameter values are injected to the relevant step executors.
Therefore, the reconfiguration can be executed without further interaction need. Thus, an administrator only needs to know \emph{what} a strategy is doing, but not \emph{how} it is realized.
	
Our current implementation supports all four aforementioned strategies to replace one EJB module with an alternative implementation of this module. The reconfiguration plan for each strategy requires only the identifiers of the replaced module and of the replacing module type as input values to perform the reconfiguration autonomically.
Nevertheless, the following restrictions must be fulfilled.

\begin{enumerate}\addtolength{\itemsep}{-0.5\baselineskip}
\item The replacing module must provide implementations for at least those interfaces that are provided by the replaced module and referenced by clients. This implies that the replacing module must fulfill the same contracts specified by these interfaces as the replaced module.
\item Each interface identified through restriction 1 must be implemented by exactly one session bean inside both, replaced and replacing modules.
\item For all required \emph{EJB References} of each of the session beans providing at least one of the interfaces identified through restriction 1, there must exist appropriate providers. An appropriate provider is a session bean which is not part of the replaced module. If the provider is part of the replacing module, this restriction must hold recursively. All \emph{EJB References} of providers not being part of the replacing module must be connected to interfaces, recursively. 
\item For each bean of the replaced module, there exists one bean in the replacing module that provides at least the same interfaces w.r.t. restriction 1.
\item For stateful session beans, the state transfer at instance level is only performed for those fields of the replaced bean - regardless of their access modifiers - for which there exists a matching counterpart in the replacing bean. In this context, two fields are matching if they have the same name and type in both, the replacing and the replaced beans.
\end{enumerate}
Though these restrictions are imposed on modules, the alternative implementation of the replacing module may eliminate failures in the behavior of the replaced one or it may be a more efficient implementation. Additionally, the integration of new functionality through adopting new or enhanced interfaces by the replacing module is possible.
\end{Section}

\begin{Section}{Conclusion and Future Work}
\label{sec:conlusion}
\noindent
With this paper we presented a flexible approach to seamless reconfiguration of EJB-based EA that need not to be anticipated during EA development, hence it maintains the idea of separation of concerns. By providing generic and reusable procedures an administrator is freed from handling	fine-grained reconfiguration tasks for each reconfiguration need. Instead of prescribing how a reconfiguration should be applied, the administrator can choose between several strategies. Thus, the role of the administrator is reduced to selecting an appropriate strategy and creating a reconfiguration plan that configures a generic procedure for a concrete reconfiguration need. The reconfiguration is performed autonomically.
	
As future work, it would be desirable if a mixture of the presented strategies could be applied for the replacement of a module, i.e., a strategy is applied only to a subset of beans of the module instead to all of its beans. Thus, disjoint subsets of beans can be reconfigured individually. Perhaps, this can be even broken down to the instance level. Finally, first considerations are made to weaken the restrictions of our current executors, e.g., to enable the replacement of $n$ modules with $m$ modules. Additionally, we investigate $x$-to-$y$ relations for the bean replacement instead of only $x$-to-$1$ relations.
\end{Section}

\bibliographystyle{latex8}
\bibliography{literature}
\end{document}